\documentclass[aps,prl,pdflatex,twocolumn,showpacs,superscriptaddress,groupedaddress]{revtex4-1}   

\usepackage{amsmath}
\usepackage{graphicx}
\usepackage{amssymb}
\usepackage[none]{hyphenat}
\usepackage{natbib}
\usepackage{color}
\usepackage{soul}

\begin{document}

\title{\bf{Measurement and shaping of biphoton spectral wavefunctions}}

\author{N. Tischler$^{1,2}$, A. B\"use$^{1,2}$, L. G. Helt$^{1,3}$, M. L. Juan$^{1,2}$, N. Piro$^{4}$, J. Ghosh$^{5}$, M. J. Steel$^{1,3}$, G. Molina-Terriza$^{1,2}$}
\email{gabriel.molina-terriza@mq.edu.au}
\affiliation{$^1$ MQ Photonics Research Centre, QSciTech, Department of Physics and Astronomy, Macquarie University, 2109, NSW, Australia}
\affiliation{$^2$ Centre of Excellence for Engineered Quantum Systems, Macquarie University, 2109, NSW, Australia}
\affiliation{$^3$ Centre of Excellence for Ultrahigh bandwidth Devices for Optical Systems, Macquarie University, 2109, NSW, Australia}
\affiliation{$^4$ \'{E}cole Polytechnique F\'ed\'erale de Lausanne (EPFL), CH-1015 Lausanne, Switzerland}
\affiliation{$^5$ Department of Physics, Indian Institute of Technology Delhi, 110016, New Delhi, India}

\date{\today}

\begin{abstract}
{In this work we present a simple method to reconstruct the complex spectral wavefunction of a biphoton, and hence gain complete information about the spectral and temporal properties of a photon pair. The technique, which relies on quantum interference, is applicable to biphoton states produced with a monochromatic pump when a shift of the pump frequency produces a shift in the relative frequencies contributing to the biphoton. We demonstrate an example of such a situation in type-II parametric down-conversion (SPDC) allowing arbitrary paraxial spatial pump and detection modes. Moreover, our test cases demonstrate the possibility to shape the spectral wavefunction. This is achieved by choosing the spatial mode of the pump and of the detection modes, and takes advantage of spatiotemporal correlations.}
\end{abstract}

\maketitle       

With the ability to exhibit nonclassical properties such as entanglement, photon pairs (or \emph{biphotons}) are of fundamental interest in quantum optics and constitute a useful resource. Many proposals exploiting this resource require knowledge of the biphoton quantum state. The complex spectral wavefunction, which contains both amplitude and phase information, provides complete knowledge of the spectrotemporal state of a photon pair, including the ability to calculate all observables related to this degree of freedom and predict interference phenomena. 
Unsurprisingly then, the reconstruction of the full complex spectral wavefunction has received a lot of attention. Several approaches have been pursued over the last years, but each proposal entails experimental challenges. Interferometric methods require a high level of stability \cite{Ren2011,Ren2012,Beduini2014}. Other methods rely on nonlinear optical effects which are inherently inefficient at the low intensity levels typical of quantum light sources, requiring very large nonlinearities or high powers \cite{Jedrkiewicz2012,ODonnell2009,Pe'er2005}. 

In the pioneering work of Hong, Ou and Mandel (HOM) \cite{Hong1987}, the coherence length and time delay between two photons was measured using quantum interference on a beam splitter, circumventing the need for optical nonlinearities. In fact, the interference phenomenon has since then proven useful in a variety of applications, including quantum teleportation \cite{Bouwmeester1997}, quantum gates \cite{Hofmann2002,Ralph2002}, linear optics quantum computation \cite{Knill2001}, Bell-state analysers \cite{Michler1996}, and the measurement of the group velocity of light \cite{Steinberg1992}, as well as of dispersion \cite{Okamoto2006}. 
Extensions of the HOM approach also enable the full reconstruction of complex spectral wavefunctions: Chen and co-workers \cite{Chen2015} rely on the time resolution of the detectors to directly measure the delay distributions, and therefore their method is applicable to very narrow-band biphotons. In contrast, Douce et al. \cite{Douce2013} propose a scheme to measure the biphoton Wigner function using HOM interference by adding shifts of the biphoton frequencies. Yet, a practical implementation of such shifts is not particularly simple or efficient \cite{Preble2012}. 

In this letter, we propose and implement a variation of the scheme in Ref.\ \cite{Douce2013} that relies on the ability to effectively shift the relative frequency of the biphoton state in the generation process. As an example, our method allows us to measure the complex spectral wavefunction, and consequently also the time delay distribution, for type-II SPDC with a monochromatic pump in an arbitrary paraxial spatial mode, after projection of the down-converted photons into a likewise arbitrary paraxial spatial mode. The assumption of a monochromatic pump beam means that the frequencies of signal and idler photons are perfectly anticorrelated, hence reducing the problem to the determination of a complex-valued function of one variable. 
Our scheme is an extension of a conventional HOM type set-up by tuning either the temperature of the nonlinear crystal or the pump frequency, so that a quantum interference coincidence pattern is recorded as a function of path length difference and crystal temperature or pump frequency. Both pump frequency tuning and crystal temperature control produce the same effect: a frequency displacement of the wavefunction. Other systems, such as four-wave mixing in atomic species can be similarly controlled by tuning the frequencies of the pumps \cite{Chen2015}. We show that multivariable quantum interference patterns can in fact be used to reconstruct the complex spectral mode function $\Phi\left(\Omega \right)$, which determines the wavefunction $|\Psi\rangle=\int\mathrm{d}\Omega~\Phi(\Omega)\hat{a}_{s}^{\dagger}\left(\frac{{\omega_{p}}}{2}+\Omega\right)\hat{a}_{i}^{\dagger}\left(\frac{{\omega_{p}}}{2}-\Omega\right)|0\rangle$. Here $\hat{a}_{m}^{\dagger}(\omega)$ is the creation operator for a photon with frequency $\omega$ and polarisation as indicated by the subscript, and $\omega_p$ is the pump frequency. 

We also demonstrate that in our chosen experimental implementation, the spectral wavefunction can be influenced through the choice of the spatial detection modes, owing to spatiotemporal correlations. This leads to nontrivial complex spectra with marked differences to the standard sinc function \cite{Fedrizzi2009}, making their characterisation worthwhile. The ability to shape the spectral wavefunction is important for quantum information and communication applications, and has already been pursued for single photons and photon pairs using other approaches \cite{Kielpinski2011,Pe'er2005}.

\begin{figure}[htbp]
\centering
\includegraphics[width=8.5cm]{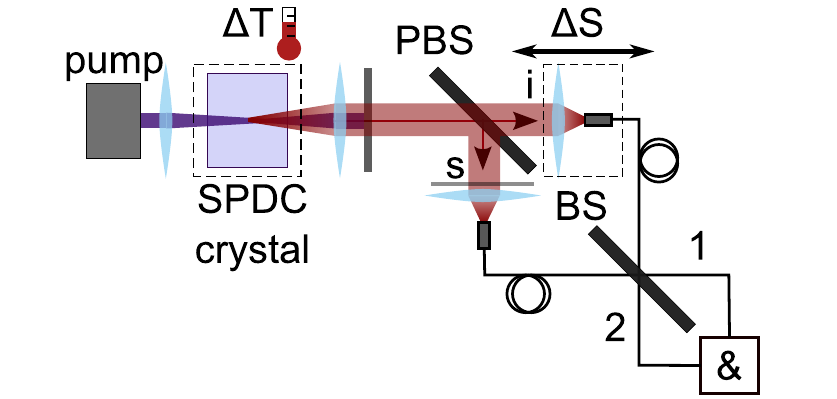}
\caption{Experimental set-up: An 8 mW, monochromatic 404.25 nm pump beam is focused to a waist of $w_p= 4.3$ $\mu$m into the temperature-controlled nonlinear crystal (15 mm, ppKTP). The down-converted light is collimated by a lens after the crystal. The pump beam is discarded by a longpass filter. The photon pairs (s and i denoting signal and idler, respectively) are separated by a polarising beam splitter (PBS). A set of waveplates (shown as the grey line in the signal path) is used to maximise interference. The path length between the two arms differs by a controllable amount $\Delta S$, before coupling into single-mode fibres. Alternatively, we can select different higher order modes with diffractive elements, prior to the fibre coupling. Finally, the photons pass through a fibre beam splitter (BS), and coincidences (\&) are detected across two avalanche photodiodes (APDs), one for each fibre beam splitter output arm.}
\label{Figure1}
\end{figure}

The experimental set-up used in the implementation of our reconstruction scheme is shown in Fig.\ \ref{Figure1}. Photon pairs are generated by pumping a periodically poled Potassium Titanyl Phosphate (ppKTP) crystal with a focused Gaussian beam in a collinear, type-II down-conversion configuration. They are then separated with a polarising beam splitter, fibre coupled, and are recombined on the two ports of a fibre-based beam splitter. As in a typical HOM experiment, the path length difference between the two arms can be swept. In addition, the temperature of the crystal is actively stabilised with a precision of $\pm$20 mK and may be tuned to any desired value. The detection mode is experimentally set to optimise the singles count rates for the chosen pump beam waist. To create different test cases, we modify the usual Gaussian detection mode by using diffractive elements to project the down-converted photons into a Laguerre-Gaussian mode, or by displacing the nonlinear crystal along the beam's propagation axis.

We measure the coincidence count rate as a joint function of path length difference and crystal temperature. Equivalently, it is possible to vary the pump laser frequency instead of the crystal temperature. We consider both cases in our theoretical analysis.

\begin{figure}[htbp]
\centering
\includegraphics[width=8.5cm]{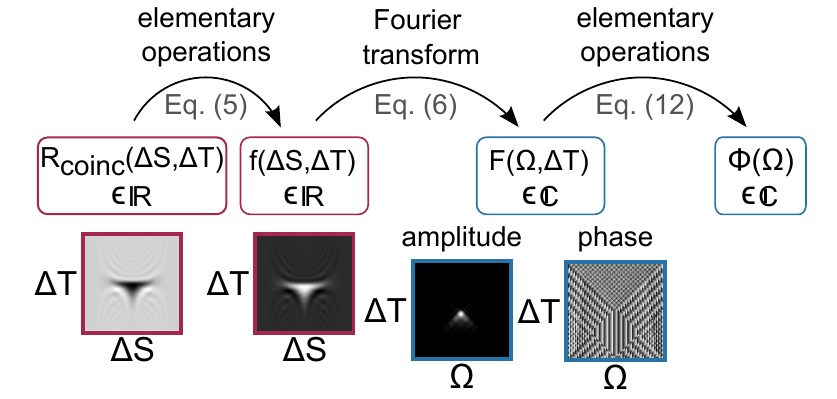}
\caption{Steps to determine $\Phi(\Omega)$ from $R_{\mathrm{coinc}}(\Delta S,\Delta T)$. Starting with $R_{\mathrm{coinc}}$, the real coincidence counts, we perform elementary operations to obtain another real function $f$ (Eq.\ (\ref{feqn})). Taking a Fourier transform of $f$ w.r.t.\ $\Delta S$, we get to the complex function $F$ (by Eq.\ (\ref{Feqn})). The desired wavefunction is obtained by taking an appropriate slice of $F$ (by Eq.\ (\ref{Formula})).}
\label{Figure2}
\end{figure}

The normalised coincidence count rates can be modelled as 
\begin{equation} 
R_{\mathrm{coinc}}(\Delta S,T,\omega_{p})= t^{4}+r^{4}-2r^{2}t^{2}\mathrm{Re}\left[f\left(\Delta S,T,\omega_{p}\right)\right],
\label{coincs}
\end{equation}
 where $\Delta S\equiv\left(S_{s}-S_{i}\right)$ is the difference between signal and idler path lengths, $T$ is the crystal temperature, $\omega_{p}$ the pump frequency, and $t$ and $r$ are the moduli of the transmission and reflection amplitudes of the HOM beam splitter, respectively. In addition, the interference term reads
\begin{eqnarray} \label{f2equation}
f\left(\Delta S,T,\omega_{p}\right)&\equiv&\int\mathrm{d}\Omega\Phi\left(\Omega;T,\omega_{p}\right)\Phi^{*}\left(-\Omega;T,\omega_{p}\right)\nonumber \\
&&\times\exp\left(i\Delta S2\Omega/\mathrm{c}\right),
\end{eqnarray}
 where $\mathrm{c}$ is the speed of light in vacuum. The conventional HOM dip is a slice of such a surface $R_{\mathrm{coinc}}(\Delta S,T,\omega_{p})$ along the $\Delta S$ direction, keeping the crystal temperature and pump frequency fixed. The measured coincidence counts thus involve our complex wavefunction of interest, 
\begin{eqnarray}
\Phi\left(\Omega;T,\omega_{p}\right)&\equiv&\int\int\mathrm{d}\mathbf{q}_{s}\mathrm{d}\mathbf{q}_{i}~\Phi_{\mathrm{full}}(\mathbf{q}_{s},\mathbf{q}_{i},\Omega;T,\omega_{p})\nonumber \\
&&\times G_{\mathrm{s}}^{*}(\mathbf{q}_{s})G_{\mathrm{i}}^{*}(\mathbf{q}_{i}),
\end{eqnarray}
where $\Phi_{\mathrm{full}}(\mathbf{q}_{s},\mathbf{q}_{i},\Omega;T,\omega_{p})$ is the wavefunction before projection into the spatial modes $G_{\mathrm{s}}(\mathbf{q}_{s})$ and $G_{\mathrm{i}}(\mathbf{q}_{i})$, with $\mathbf{q}$ being the transverse momenta  \footnote{Details can be found in the supplementary information, which includes Refs.\ \cite{Buese2015,Kato2002,Pignatiello2007,Ghosh2014}}. \phantom{\cite{Buese2015,Kato2002,Pignatiello2007,Ghosh2014}} However, due to the nature of quantum interference, the wavefunction appears in the form of 
\begin{equation}
F(\Omega,T,\omega_{p})\equiv\Phi\left(\Omega;T,\omega_{p}\right)\Phi^{*}\left(-\Omega;T,\omega_{p}\right),
\label{sym}
\end{equation}
which we refer to as the \emph{symmetrised} wavefunction. 
Since $F(\Omega,T,\omega_{p})$ is Hermitian w.r.t.\ $\Omega$, $f\left(\Delta S,T,\omega_{p}\right)$ is real and from the coincidence rates (recall Eq.\ (\ref{coincs})):
\begin{equation}\label{feqn} 
f\left(\Delta S,T,\omega_{p}\right)=\left(\frac{1}{2r^{2}t^{2}}\left(t^{4}+r^{4} -R_{\mathrm{coinc}}(\Delta S,T,\omega_{p})\right)\right).
\end{equation}
We then obtain $F(\Omega,T,\omega_{p})$ by taking the Fourier transform of $f\left(\Delta S,T,\omega_{p}\right)$ with respect to $\Delta S$:
\begin{equation}\label{Feqn} 
 F(\Omega, T, \omega_{p})=\frac{1}{c\pi}\int\mathrm{d}\Delta Sf\left(\Delta S,T,\omega_{p}\right)  \mathrm{exp}(-i(2\Omega)\Delta S/c). 
\end{equation}
Because the symmetrisation is not isomorphic, Eq.\ (\ref{sym}) can in general not be inverted to retrieve the wavefunction from the usual HOM dip, so additional information is required. One possibility is to extend the measurements by shifting the relative frequencies of signal and idler. We now show that the reconstruction is also possible by performing a temperature or a pump frequency sweep. To see this, we perform a multivariate Taylor expansion to leading orders of the wavevector z-components for the pump, signal, and idler (indicated by subscript $m$), about the values at which perfect phase matching takes place: at frequencies $\omega_{m}=\omega_{0m}$, crystal temperature $T=T_{0}$, and transverse wavevector $\mathbf{q}_{m}=\mathbf{0}$. From the Taylor series approximation, symmetrised wavefunctions at different temperatures can be related by shifting the frequencies, while keeping the temperature fixed \footnotemark[\value{footnote}]:
\begin{eqnarray}\label{MethodCondition}
 &\Phi&\left(\Omega;T_{0}+\Delta T,\omega_{0p}+\Delta\omega_{p}\right)\Phi^{*}\left(-\Omega;T_{0}+\Delta T,\omega_{0p}+\Delta\omega_{p}\right) \nonumber \\
&\approx&\Phi\left(\Omega+\Delta Tc_{t}+\Delta\omega_{p}c_{\omega p};T_{0},\omega_{0p}\right)\nonumber \\
&&\times \Phi^{*}\left(-\Omega+\Delta Tc_{t}+\Delta\omega_{p}c_{\omega p};T_{0},\omega_{0p}\right),
\end{eqnarray}
where we have defined
\begin{eqnarray} 
c_{t}&\equiv&-\frac{{X_{T}}}{\left(\frac{{\partial k_{s}}}{\partial\omega}-\frac{{\partial k_{i}}}{\partial\omega}\right)}, \\
 X_{T}&\equiv&\left(\frac{{\partial k_{p}}}{\partial T}-\frac{{\partial k_{s}}}{\partial T}-\frac{{\partial k_{i}}}{\partial T}+\frac{2\pi}{\left(\Lambda\left(T_{0}\right)\right)^{2}}\frac{\partial\Lambda}{\partial T}\right),\\
 c_{\omega p}&\equiv&-\frac{{X_{\omega}}}{\left(\frac{{\partial k_{s}}}{\partial\omega}-\frac{{\partial k_{i}}}{\partial\omega}\right)},\\
X_{\omega}&\equiv&\left(\frac{{\partial k_{p}}}{\partial\omega}-\frac{{\partial k_{s}}}{2\partial\omega}-\frac{{\partial k_{i}}}{2\partial\omega}\right).
\end{eqnarray}
Here, $k_m$ are the wavenumbers in the crystal (pump, signal, and idler indicated by subscripts), $\omega_m$ the frequencies, and $\Lambda$ the poling period of the crystal. The $\frac{\partial k_m}{\partial\omega}$ are inverse group velocities. All derivatives are evaluated at the reference temperature $T_{0}$ and frequencies $\omega_{m}=\omega_{0m}$, $m\in \{p,s,i \}$. We identify $c_t$ and $c_{\omega p}$ as proportionality constants between a shift in $T$ or $\omega_p$, and $\Omega$. They represent the measurable sensitivity of the biphoton spectrum to the crystal temperature and pump wavelength. The complex mode function can be obtained as a slice through $F(\Omega,\Delta T,\Delta \omega_{p})$ : 
\begin{eqnarray}\label{Formula}
&\Phi&(2c_{t}\Delta T+2c_{\omega p}\Delta\omega_{p};T_{0},\omega_{0p})\nonumber \\
&=&\frac{{F^*(-c_{t}\Delta T-c_{\omega p}\Delta\omega_{p},T_{0}+\Delta T,\omega_{0p}+\Delta\omega_{p})}}{\sqrt{|{F(0,T_{0},\omega_{0p})|}}}.\nonumber \\
\end{eqnarray}
Fig.\ \ref{Figure2} illustrates the data analysis process, when we choose to do a sweep in the temperature.

\begin{figure*}[htp]
\centering
\includegraphics[width=17cm]{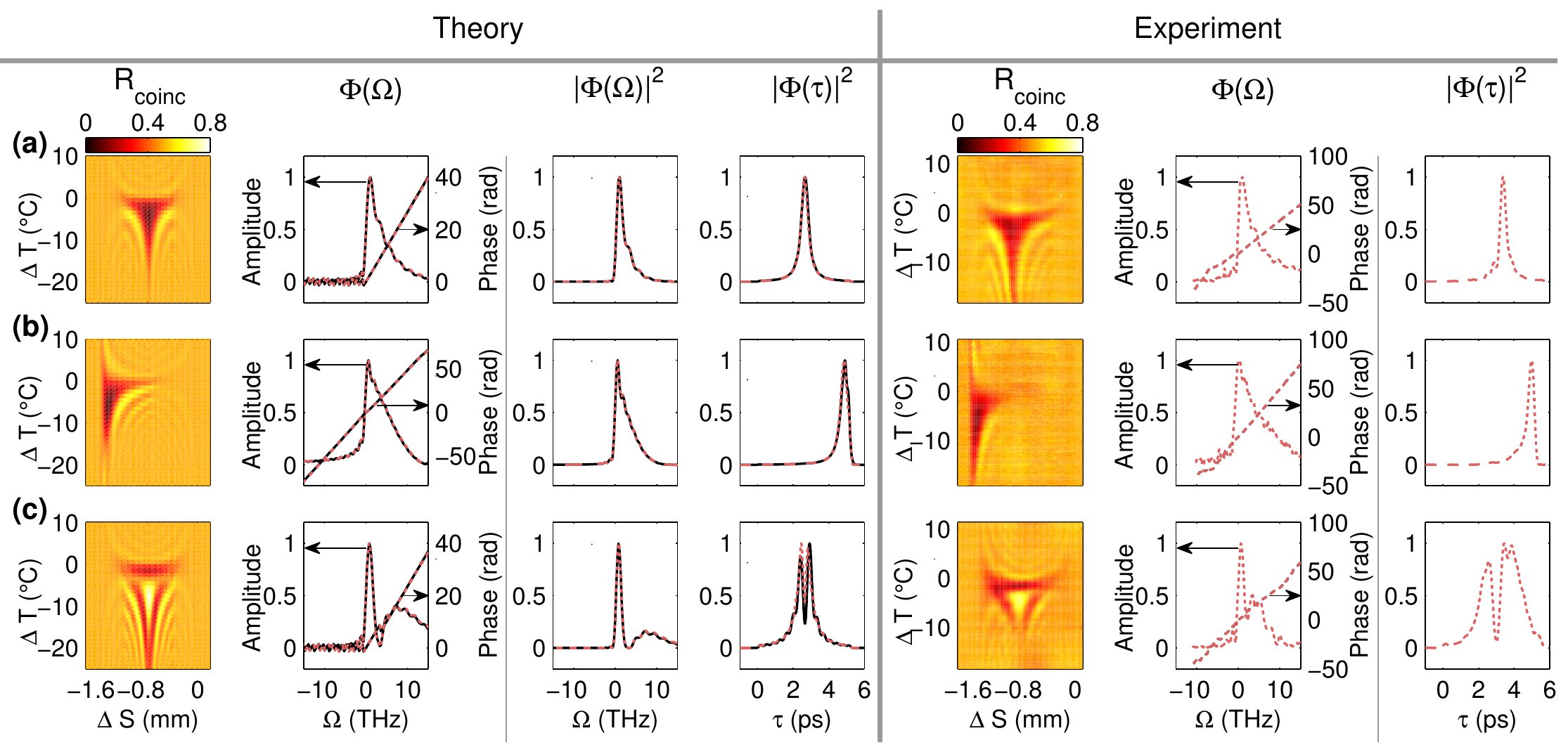}
\caption{Demonstration of the reconstruction method on three test cases. Theoretical (left) and experimental (right panel) results using as detection modes (a) Gaussians with the crystal centered, (b) Gaussians with the crystal displaced by 3 mm along the propagation direction, and (c) a pair of Laguerre Gaussian modes (1 0), (-1 0) with the crystal centered. We show, in both cases, (from left to right) the coincidence counts, the complex spectral wavefunctions (amplitude in arb.\ units), the spectral distributions (arb.\ units), and the time delay distributions (arb.\ units). For experiments, we omit the spectral distribution. The red dotted lines are the reconstruction results, while the black solid lines within the theory section are based on the simulated wavefunction. Our experimental conditions are simulated using $T_0=58$ $^\circ$C, so that $c_t=-4.8698 \times 10^{11}$ ($^\circ$C$\cdot$s)$^{-1}$ and with a detection beam waist of 9.6 $\mu m$.}
\label{Figure3}
\end{figure*}

In Fig.\ \ref{Figure3} we demonstrate the use of our reconstruction method, with theory and experimental results for three test cases. The test cases all use a Gaussian pump beam with a beam waist of 4.3 $\mu \mathrm{m}$, but differ in the detection modes. These are (a) Gaussians with the crystal centered, (b) Gaussians with the crystal displaced by 3 mm along the propagation direction, and (c) the Laguerre Gaussian modes (azimuthal index, radial index) = (1,0), (-1,0) with the crystal centered.

We first show the results of our theoretical analysis, illustrated in the left panel of Fig.\ 3. From our model of the nonlinear process and detection, we obtain the expected spectral wavefunction directly (black solid line) \footnotemark[\value{footnote}]. We then calculate the expected quantum interference pattern, $R_{\mathrm{coinc}}$, by Eqs.\ (\ref{coincs}) and (\ref{f2equation}), based on which the spectral wavefunction is reconstructed using Eqs.\ (\ref{feqn}), (\ref{Feqn}) and (\ref{Formula}) (red dotted line). Next we calculate the spectral and time delay distributions both from the original and the reconstructed wavefunction, shown in black solid and red dotted lines, respectively. 
For the experimental results, we measure the quantum interference pattern and reconstruct the wavefunction using Eqs.\ (\ref{feqn}), (\ref{Feqn}) and (\ref{Formula}), from which the time delay distribution is obtained (red dotted line).

The theoretical analysis allows us to compare the original and the reconstructed wavefunction, showing a good agreement in all cases. However, the reconstruction is insensitive to those quadratic and higher order phases as a function of $\Omega$ that stem from the propagation of the photon pairs to the end of the crystal, or through any additional dispersive elements. This causes an error in the reconstructed phase \footnotemark[\value{footnote}]. For any particular implementation, the error is limited and depends on the optical properties and length of the crystal, as well as the spectral bandwidth. The spectral bandwidth is influenced by the detection mode, and our choice of a small detection beam waist corresponds to a broad spectrum, which allows us to explore limitations of the method. The deviation in phase cannot be seen easily in Fig.\ \ref{Figure3} because it is comparatively small, but there is a difference which is quadratic in $\Omega$ and reaches up to 0.46 radians for the plotted section of frequencies. This, in turn, has a visible impact on the time delay distribution in (c), where a small deviation between reconstructed and calculated distributions is evident. We note that this lack of sensitivity of the reconstruction method does not mean that it is overall only sensitive to linear functions of the phase, as the phase imparted through the spatial projection can be arbitrary and is recovered by our method. 

The experimentally measured coincidences allow us to determine the complex spectral wavefunctions and time delay distributions of our experimental photon pairs. Interestingly, they also allow us to identify small imperfections in the experiment. For example, a slight off-centering of the crystal in cases (a) and (c) results in an asymmetry of the coincidence count map w.r.t.\ $\Delta S$ and an increased slope of the phase and mean time delay. In (c), we attribute differences with the theoretically predicted wavefunction to the fact that the radial profile of the theoretical detection mode is slightly different from the one in the experimental implementation.

A comparison of the three rows in Fig.\ \ref{Figure3} shows significant differences between the test cases. The Gaussian detection with the crystal centred (a) yields a quantum interference pattern that is symmetric w.r.t.\ $\Delta S$, about a value that depends on the time delay acquired when signal and idler photons propagate through half the length of the nonlinear birefringent crystal. The spectrum's departure from a sinc squared function is highlighted by its asymmetry. It arises from our use of a small detection beam waist, and also results in the asymmetry of the quantum interference pattern in the $\Delta T$ direction. The time delay distribution has a symmetric peak centred at the time delay acquired by propagation through half of the crystal. When the crystal is displaced (b), the quantum interference pattern becomes asymmetric, the phase changes, and the time delay distribution shifts \cite{Buese2015}. Using the Laguerre Gaussian detection mode with the crystal centred (c) changes the structure of the quantum interference pattern markedly, even transforming the dip into a peak. The phase of the wavefunction is similar to the Gaussian case, but the spectrum has a side lobe. Interestingly, the time delay distribution has a dip at the approximate axis of symmetry, which means that the probability of photons arriving with their mean time delay is suppressed. 

In summary, we have proposed and demonstrated a method to reconstruct the complex spectral wavefunction of a biphoton, using HOM interference for type-II SPDC. The essence of our method lies in the fact that a change in temperature or pump frequency is approximately equivalent to a shift of the frequency for the symmetrised mode function that determines the quantum interference coincidence counts. 

A considerable advantage of the method lies in its simplicity, both in the experimental implementation and in the data analysis. 
Indeed, our technique is not faced with challenging stabilisations typical of interferometric measurements \cite{Beduini2014}, or the need for high pump powers  incurred by measurements that rely on nonlinear optical effects \cite{ODonnell2009}. As an extension, following the results in Ref.\ \cite{Douce2013}, $f\left(\Delta S,T,\omega_{p}\right)$ provides the Wigner function in the case of a mixed state. For the case of a pulsed pump where the biphoton wavefunction depends on both signal and idler frequencies, $f\left(\Delta S,T,\omega_{0p}\right)$ can provide the Wigner function in which the sum frequency variable has been traced out. We note that to appropriately manipulate the biphotons in the generation process, so that their wavefunction can be shifted as shown here, the argument of the phase matching function must be linear in the relative frequency $\Omega$. This includes processes such as collinear and noncollinear type-II SPDC with or without periodic poling \footnotemark[\value{footnote}], but not type-I collinear degenerate down-conversion due to the group velocities of the two photons being equal. Moreover, post-emission spectral manipulation, such as the use of spectral filters or propagation through dispersive elements, will lead to a faulty reconstruction. 
A further limitation is that the reconstruction is unsuccessful at recovering the limited part of the quadratic and higher order phase that arises from the propagation of the biphoton to the end of the crystal \footnotemark[\value{footnote}].

Lastly, the freedom to choose spatial pump and detection modes offers some interesting possibilities. We have characterised the spectrotemporal properties of the biphoton, after projection into a spatial mode. Contrary to the intuitive idea that spatial degrees of freedom should not play a role, our results show that the choice of detection modes can have a pronounced effect on the spectrotemporal properties, in particular due to spatiotemporal correlations in the biphoton wavefunction \cite{Brambilla2010,Osorio2008,Gatti2009,Buese2015}. Our method works for arbitrary paraxial pump and projection modes, so it is possible to influence the detected wavefunction by adjusting the modes.\\

This work was funded by the Australian Research Council's Centres of Excellence for Engineered Quantum Systems (EQuS), grant number CE110001013, and for Ultrahigh bandwidth Devices for Optical Systems (CUDOS), grant number CE110001018. G.M.-T. also holds an Australian Research Council Future Fellowship.

\bibliography{WFRpaper_references2}

\begin{thebibliography}{27}%
\makeatletter
\providecommand \@ifxundefined [1]{%
 \@ifx{#1\undefined}
}%
\providecommand \@ifnum [1]{%
 \ifnum #1\expandafter \@firstoftwo
 \else \expandafter \@secondoftwo
 \fi
}%
\providecommand \@ifx [1]{%
 \ifx #1\expandafter \@firstoftwo
 \else \expandafter \@secondoftwo
 \fi
}%
\providecommand \natexlab [1]{#1}%
\providecommand \enquote  [1]{``#1''}%
\providecommand \bibnamefont  [1]{#1}%
\providecommand \bibfnamefont [1]{#1}%
\providecommand \citenamefont [1]{#1}%
\providecommand \href@noop [0]{\@secondoftwo}%
\providecommand \href [0]{\begingroup \@sanitize@url \@href}%
\providecommand \@href[1]{\@@startlink{#1}\@@href}%
\providecommand \@@href[1]{\endgroup#1\@@endlink}%
\providecommand \@sanitize@url [0]{\catcode `\\12\catcode `\$12\catcode
  `\&12\catcode `\#12\catcode `\^12\catcode `\_12\catcode `\%12\relax}%
\providecommand \@@startlink[1]{}%
\providecommand \@@endlink[0]{}%
\providecommand \url  [0]{\begingroup\@sanitize@url \@url }%
\providecommand \@url [1]{\endgroup\@href {#1}{\urlprefix }}%
\providecommand \urlprefix  [0]{URL }%
\providecommand \Eprint [0]{\href }%
\providecommand \doibase [0]{http://dx.doi.org/}%
\providecommand \selectlanguage [0]{\@gobble}%
\providecommand \bibinfo  [0]{\@secondoftwo}%
\providecommand \bibfield  [0]{\@secondoftwo}%
\providecommand \translation [1]{[#1]}%
\providecommand \BibitemOpen [0]{}%
\providecommand \bibitemStop [0]{}%
\providecommand \bibitemNoStop [0]{.\EOS\space}%
\providecommand \EOS [0]{\spacefactor3000\relax}%
\providecommand \BibitemShut  [1]{\csname bibitem#1\endcsname}%
\let\auto@bib@innerbib\@empty
\bibitem [{\citenamefont {Ren}\ and\ \citenamefont {Hofmann}(2011)}]{Ren2011}%
  \BibitemOpen
  \bibfield  {author} {\bibinfo {author} {\bibfnamefont {C.}~\bibnamefont
  {Ren}}\ and\ \bibinfo {author} {\bibfnamefont {H.~F.}\ \bibnamefont
  {Hofmann}},\ }\href {\doibase 10.1103/PhysRevA.84.032108} {\bibfield
  {journal} {\bibinfo  {journal} {Physical Review A}\ }\textbf {\bibinfo
  {volume} {84}},\ \bibinfo {pages} {032108} (\bibinfo {year}
  {2011})}\BibitemShut {NoStop}%
\bibitem [{\citenamefont {Ren}\ and\ \citenamefont {Hofmann}(2012)}]{Ren2012}%
  \BibitemOpen
  \bibfield  {author} {\bibinfo {author} {\bibfnamefont {C.}~\bibnamefont
  {Ren}}\ and\ \bibinfo {author} {\bibfnamefont {H.~F.}\ \bibnamefont
  {Hofmann}},\ }\href {\doibase 10.1103/PhysRevA.86.043823} {\bibfield
  {journal} {\bibinfo  {journal} {Physical Review A}\ }\textbf {\bibinfo
  {volume} {86}},\ \bibinfo {pages} {043823} (\bibinfo {year}
  {2012})}\BibitemShut {NoStop}%
\bibitem [{\citenamefont {Beduini}\ \emph {et~al.}(2014)\citenamefont
  {Beduini}, \citenamefont {Zielinska}, \citenamefont {Lucivero}, \citenamefont
  {{de Icaza Astiz}},\ and\ \citenamefont {Mitchell}}]{Beduini2014}%
  \BibitemOpen
  \bibfield  {author} {\bibinfo {author} {\bibfnamefont {F.~A.}\ \bibnamefont
  {Beduini}}, \bibinfo {author} {\bibfnamefont {J.~A.}\ \bibnamefont
  {Zielinska}}, \bibinfo {author} {\bibfnamefont {V.~G.}\ \bibnamefont
  {Lucivero}}, \bibinfo {author} {\bibfnamefont {Y.~A.}\ \bibnamefont {{de
  Icaza Astiz}}}, \ and\ \bibinfo {author} {\bibfnamefont {M.~W.}\ \bibnamefont
  {Mitchell}},\ }\href {\doibase 10.1103/PhysRevLett.113.183602} {\bibfield
  {journal} {\bibinfo  {journal} {Physical Review Letters}\ }\textbf {\bibinfo
  {volume} {113}},\ \bibinfo {pages} {183602} (\bibinfo {year}
  {2014})}\BibitemShut {NoStop}%
\bibitem [{\citenamefont {Jedrkiewicz}\ \emph {et~al.}(2012)\citenamefont
  {Jedrkiewicz}, \citenamefont {Blanchet}, \citenamefont {Brambilla},
  \citenamefont {{Di Trapani}},\ and\ \citenamefont {Gatti}}]{Jedrkiewicz2012}%
  \BibitemOpen
  \bibfield  {author} {\bibinfo {author} {\bibfnamefont {O.}~\bibnamefont
  {Jedrkiewicz}}, \bibinfo {author} {\bibfnamefont {J.-L.}\ \bibnamefont
  {Blanchet}}, \bibinfo {author} {\bibfnamefont {E.}~\bibnamefont {Brambilla}},
  \bibinfo {author} {\bibfnamefont {P.}~\bibnamefont {{Di Trapani}}}, \ and\
  \bibinfo {author} {\bibfnamefont {A.}~\bibnamefont {Gatti}},\ }\href
  {\doibase 10.1103/PhysRevLett.108.253904} {\bibfield  {journal} {\bibinfo
  {journal} {Physical Review Letters}\ }\textbf {\bibinfo {volume} {108}},\
  \bibinfo {pages} {253904} (\bibinfo {year} {2012})}\BibitemShut {NoStop}%
\bibitem [{\citenamefont {O'Donnell}\ and\ \citenamefont
  {U'Ren}(2009)}]{ODonnell2009}%
  \BibitemOpen
  \bibfield  {author} {\bibinfo {author} {\bibfnamefont {K.}~\bibnamefont
  {O'Donnell}}\ and\ \bibinfo {author} {\bibfnamefont {A.}~\bibnamefont
  {U'Ren}},\ }\href {\doibase 10.1103/PhysRevLett.103.123602} {\bibfield
  {journal} {\bibinfo  {journal} {Physical Review Letters}\ }\textbf {\bibinfo
  {volume} {103}},\ \bibinfo {pages} {123602} (\bibinfo {year}
  {2009})}\BibitemShut {NoStop}%
\bibitem [{\citenamefont {Pe'er}\ \emph {et~al.}(2005)\citenamefont {Pe'er},
  \citenamefont {Dayan}, \citenamefont {Friesem},\ and\ \citenamefont
  {Silberberg}}]{Pe'er2005}%
  \BibitemOpen
  \bibfield  {author} {\bibinfo {author} {\bibfnamefont {A.}~\bibnamefont
  {Pe'er}}, \bibinfo {author} {\bibfnamefont {B.}~\bibnamefont {Dayan}},
  \bibinfo {author} {\bibfnamefont {A.}~\bibnamefont {Friesem}}, \ and\
  \bibinfo {author} {\bibfnamefont {Y.}~\bibnamefont {Silberberg}},\ }\href
  {\doibase 10.1103/PhysRevLett.94.073601} {\bibfield  {journal} {\bibinfo
  {journal} {Physical Review Letters}\ }\textbf {\bibinfo {volume} {94}},\
  \bibinfo {pages} {073601} (\bibinfo {year} {2005})}\BibitemShut {NoStop}%
\bibitem [{\citenamefont {Hong}\ \emph {et~al.}(1987)\citenamefont {Hong},
  \citenamefont {Ou},\ and\ \citenamefont {Mandel}}]{Hong1987}%
  \BibitemOpen
  \bibfield  {author} {\bibinfo {author} {\bibfnamefont {C.~K.}\ \bibnamefont
  {Hong}}, \bibinfo {author} {\bibfnamefont {Z.~Y.}\ \bibnamefont {Ou}}, \ and\
  \bibinfo {author} {\bibfnamefont {L.}~\bibnamefont {Mandel}},\ }\href
  {\doibase 10.1103/PhysRevLett.59.2044} {\bibfield  {journal} {\bibinfo
  {journal} {Physical Review Letters}\ }\textbf {\bibinfo {volume} {59}},\
  \bibinfo {pages} {2044} (\bibinfo {year} {1987})}\BibitemShut {NoStop}%
\bibitem [{\citenamefont {Bouwmeester}\ \emph {et~al.}(1997)\citenamefont
  {Bouwmeester}, \citenamefont {Pan}, \citenamefont {Mattle}, \citenamefont
  {Eibl}, \citenamefont {Weinfurter},\ and\ \citenamefont
  {Zeilinger}}]{Bouwmeester1997}%
  \BibitemOpen
  \bibfield  {author} {\bibinfo {author} {\bibfnamefont {D.}~\bibnamefont
  {Bouwmeester}}, \bibinfo {author} {\bibfnamefont {J.-W.}\ \bibnamefont
  {Pan}}, \bibinfo {author} {\bibfnamefont {K.}~\bibnamefont {Mattle}},
  \bibinfo {author} {\bibfnamefont {M.}~\bibnamefont {Eibl}}, \bibinfo {author}
  {\bibfnamefont {H.}~\bibnamefont {Weinfurter}}, \ and\ \bibinfo {author}
  {\bibfnamefont {A.}~\bibnamefont {Zeilinger}},\ }\href {\doibase
  10.1038/37539} {\bibfield  {journal} {\bibinfo  {journal} {Nature}\ }\textbf
  {\bibinfo {volume} {390}},\ \bibinfo {pages} {575} (\bibinfo {year}
  {1997})}\BibitemShut {NoStop}%
\bibitem [{\citenamefont {Hofmann}\ and\ \citenamefont
  {Takeuchi}(2002)}]{Hofmann2002}%
  \BibitemOpen
  \bibfield  {author} {\bibinfo {author} {\bibfnamefont {H.}~\bibnamefont
  {Hofmann}}\ and\ \bibinfo {author} {\bibfnamefont {S.}~\bibnamefont
  {Takeuchi}},\ }\href {\doibase 10.1103/PhysRevA.66.024308} {\bibfield
  {journal} {\bibinfo  {journal} {Physical Review A}\ }\textbf {\bibinfo
  {volume} {66}},\ \bibinfo {pages} {024308} (\bibinfo {year}
  {2002})}\BibitemShut {NoStop}%
\bibitem [{\citenamefont {Ralph}\ \emph {et~al.}(2002)\citenamefont {Ralph},
  \citenamefont {Langford}, \citenamefont {Bell},\ and\ \citenamefont
  {White}}]{Ralph2002}%
  \BibitemOpen
  \bibfield  {author} {\bibinfo {author} {\bibfnamefont {T.}~\bibnamefont
  {Ralph}}, \bibinfo {author} {\bibfnamefont {N.}~\bibnamefont {Langford}},
  \bibinfo {author} {\bibfnamefont {T.}~\bibnamefont {Bell}}, \ and\ \bibinfo
  {author} {\bibfnamefont {A.}~\bibnamefont {White}},\ }\href {\doibase
  10.1103/PhysRevA.65.062324} {\bibfield  {journal} {\bibinfo  {journal}
  {Physical Review A}\ }\textbf {\bibinfo {volume} {65}},\ \bibinfo {pages}
  {062324} (\bibinfo {year} {2002})}\BibitemShut {NoStop}%
\bibitem [{\citenamefont {Knill}\ \emph {et~al.}(2001)\citenamefont {Knill},
  \citenamefont {Laflamme},\ and\ \citenamefont {Milburn}}]{Knill2001}%
  \BibitemOpen
  \bibfield  {author} {\bibinfo {author} {\bibfnamefont {E.}~\bibnamefont
  {Knill}}, \bibinfo {author} {\bibfnamefont {R.}~\bibnamefont {Laflamme}}, \
  and\ \bibinfo {author} {\bibfnamefont {G.~J.}\ \bibnamefont {Milburn}},\
  }\href {\doibase 10.1038/35051009} {\bibfield  {journal} {\bibinfo  {journal}
  {Nature}\ }\textbf {\bibinfo {volume} {409}},\ \bibinfo {pages} {46}
  (\bibinfo {year} {2001})}\BibitemShut {NoStop}%
\bibitem [{\citenamefont {Michler}\ \emph {et~al.}(1996)\citenamefont
  {Michler}, \citenamefont {Mattle}, \citenamefont {Weinfurter},\ and\
  \citenamefont {Zeilinger}}]{Michler1996}%
  \BibitemOpen
  \bibfield  {author} {\bibinfo {author} {\bibfnamefont {M.}~\bibnamefont
  {Michler}}, \bibinfo {author} {\bibfnamefont {K.}~\bibnamefont {Mattle}},
  \bibinfo {author} {\bibfnamefont {H.}~\bibnamefont {Weinfurter}}, \ and\
  \bibinfo {author} {\bibfnamefont {A.}~\bibnamefont {Zeilinger}},\ }\href
  {http://www.ncbi.nlm.nih.gov/pubmed/9913093} {\bibfield  {journal} {\bibinfo
  {journal} {Physical Review A}\ }\textbf {\bibinfo {volume} {53}},\ \bibinfo
  {pages} {R1209} (\bibinfo {year} {1996})}\BibitemShut {NoStop}%
\bibitem [{\citenamefont {Steinberg}\ \emph {et~al.}(1992)\citenamefont
  {Steinberg}, \citenamefont {Kwiat},\ and\ \citenamefont
  {Chiao}}]{Steinberg1992}%
  \BibitemOpen
  \bibfield  {author} {\bibinfo {author} {\bibfnamefont {A.}~\bibnamefont
  {Steinberg}}, \bibinfo {author} {\bibfnamefont {P.}~\bibnamefont {Kwiat}}, \
  and\ \bibinfo {author} {\bibfnamefont {R.}~\bibnamefont {Chiao}},\ }\href
  {\doibase 10.1103/PhysRevLett.68.2421} {\bibfield  {journal} {\bibinfo
  {journal} {Physical Review Letters}\ }\textbf {\bibinfo {volume} {68}},\
  \bibinfo {pages} {2421} (\bibinfo {year} {1992})}\BibitemShut {NoStop}%
\bibitem [{\citenamefont {Okamoto}\ \emph {et~al.}(2006)\citenamefont
  {Okamoto}, \citenamefont {Takeuchi},\ and\ \citenamefont
  {Sasaki}}]{Okamoto2006}%
  \BibitemOpen
  \bibfield  {author} {\bibinfo {author} {\bibfnamefont {R.}~\bibnamefont
  {Okamoto}}, \bibinfo {author} {\bibfnamefont {S.}~\bibnamefont {Takeuchi}}, \
  and\ \bibinfo {author} {\bibfnamefont {K.}~\bibnamefont {Sasaki}},\ }\href
  {\doibase 10.1103/PhysRevA.74.011801} {\bibfield  {journal} {\bibinfo
  {journal} {Physical Review A}\ }\textbf {\bibinfo {volume} {74}},\ \bibinfo
  {pages} {011801} (\bibinfo {year} {2006})}\BibitemShut {NoStop}%
\bibitem [{\citenamefont {Chen}\ \emph {et~al.}(2015)\citenamefont {Chen},
  \citenamefont {Shu}, \citenamefont {Guo}, \citenamefont {Loy},\ and\
  \citenamefont {Du}}]{Chen2015}%
  \BibitemOpen
  \bibfield  {author} {\bibinfo {author} {\bibfnamefont {P.}~\bibnamefont
  {Chen}}, \bibinfo {author} {\bibfnamefont {C.}~\bibnamefont {Shu}}, \bibinfo
  {author} {\bibfnamefont {X.}~\bibnamefont {Guo}}, \bibinfo {author}
  {\bibfnamefont {M.}~\bibnamefont {Loy}}, \ and\ \bibinfo {author}
  {\bibfnamefont {S.}~\bibnamefont {Du}},\ }\href {\doibase
  10.1103/PhysRevLett.114.010401} {\bibfield  {journal} {\bibinfo  {journal}
  {Physical Review Letters}\ }\textbf {\bibinfo {volume} {114}},\ \bibinfo
  {pages} {010401} (\bibinfo {year} {2015})}\BibitemShut {NoStop}%
\bibitem [{\citenamefont {Douce}\ \emph {et~al.}(2013)\citenamefont {Douce},
  \citenamefont {Eckstein}, \citenamefont {Walborn}, \citenamefont {Khoury},
  \citenamefont {Ducci}, \citenamefont {Keller}, \citenamefont {Coudreau},\
  and\ \citenamefont {Milman}}]{Douce2013}%
  \BibitemOpen
  \bibfield  {author} {\bibinfo {author} {\bibfnamefont {T.}~\bibnamefont
  {Douce}}, \bibinfo {author} {\bibfnamefont {A.}~\bibnamefont {Eckstein}},
  \bibinfo {author} {\bibfnamefont {S.~P.}\ \bibnamefont {Walborn}}, \bibinfo
  {author} {\bibfnamefont {A.~Z.}\ \bibnamefont {Khoury}}, \bibinfo {author}
  {\bibfnamefont {S.}~\bibnamefont {Ducci}}, \bibinfo {author} {\bibfnamefont
  {A.}~\bibnamefont {Keller}}, \bibinfo {author} {\bibfnamefont
  {T.}~\bibnamefont {Coudreau}}, \ and\ \bibinfo {author} {\bibfnamefont
  {P.}~\bibnamefont {Milman}},\ }\href {\doibase 10.1038/srep03530} {\bibfield
  {journal} {\bibinfo  {journal} {Scientific reports}\ }\textbf {\bibinfo
  {volume} {3}},\ \bibinfo {pages} {3530} (\bibinfo {year} {2013})}\BibitemShut
  {NoStop}%
\bibitem [{\citenamefont {Preble}\ \emph {et~al.}(2012)\citenamefont {Preble},
  \citenamefont {Cao}, \citenamefont {Elshaari}, \citenamefont {Aboketaf},\
  and\ \citenamefont {Adams}}]{Preble2012}%
  \BibitemOpen
  \bibfield  {author} {\bibinfo {author} {\bibfnamefont {S.}~\bibnamefont
  {Preble}}, \bibinfo {author} {\bibfnamefont {L.}~\bibnamefont {Cao}},
  \bibinfo {author} {\bibfnamefont {A.}~\bibnamefont {Elshaari}}, \bibinfo
  {author} {\bibfnamefont {A.}~\bibnamefont {Aboketaf}}, \ and\ \bibinfo
  {author} {\bibfnamefont {D.}~\bibnamefont {Adams}},\ }\href {\doibase
  10.1063/1.4764068} {\bibfield  {journal} {\bibinfo  {journal} {Applied
  Physics Letters}\ }\textbf {\bibinfo {volume} {101}},\ \bibinfo {pages}
  {171110} (\bibinfo {year} {2012})}\BibitemShut {NoStop}%
\bibitem [{\citenamefont {Fedrizzi}\ \emph {et~al.}(2009)\citenamefont
  {Fedrizzi}, \citenamefont {Herbst}, \citenamefont {Aspelmeyer}, \citenamefont
  {Barbieri}, \citenamefont {Jennewein},\ and\ \citenamefont
  {Zeilinger}}]{Fedrizzi2009}%
  \BibitemOpen
  \bibfield  {author} {\bibinfo {author} {\bibfnamefont {A.}~\bibnamefont
  {Fedrizzi}}, \bibinfo {author} {\bibfnamefont {T.}~\bibnamefont {Herbst}},
  \bibinfo {author} {\bibfnamefont {M.}~\bibnamefont {Aspelmeyer}}, \bibinfo
  {author} {\bibfnamefont {M.}~\bibnamefont {Barbieri}}, \bibinfo {author}
  {\bibfnamefont {T.}~\bibnamefont {Jennewein}}, \ and\ \bibinfo {author}
  {\bibfnamefont {A.}~\bibnamefont {Zeilinger}},\ }\href {\doibase
  10.1088/1367-2630/11/10/103052} {\bibfield  {journal} {\bibinfo  {journal}
  {New Journal of Physics}\ }\textbf {\bibinfo {volume} {11}},\ \bibinfo
  {pages} {103052} (\bibinfo {year} {2009})}\BibitemShut {NoStop}%
\bibitem [{\citenamefont {Kielpinski}\ \emph {et~al.}(2011)\citenamefont
  {Kielpinski}, \citenamefont {Corney},\ and\ \citenamefont
  {Wiseman}}]{Kielpinski2011}%
  \BibitemOpen
  \bibfield  {author} {\bibinfo {author} {\bibfnamefont {D.}~\bibnamefont
  {Kielpinski}}, \bibinfo {author} {\bibfnamefont {J.~F.}\ \bibnamefont
  {Corney}}, \ and\ \bibinfo {author} {\bibfnamefont {H.~M.}\ \bibnamefont
  {Wiseman}},\ }\href {\doibase 10.1103/PhysRevLett.106.130501} {\bibfield
  {journal} {\bibinfo  {journal} {Physical Review Letters}\ }\textbf {\bibinfo
  {volume} {106}},\ \bibinfo {pages} {130501} (\bibinfo {year}
  {2011})}\BibitemShut {NoStop}%
\bibitem [{Note1()}]{Note1}%
  \BibitemOpen
  \bibinfo {note} {Details can be found in the supplementary information, which
  includes Refs.\ \cite
  {Buese2015,Kato2002,Pignatiello2007,Ghosh2014}}\BibitemShut {NoStop}%
\bibitem [{\citenamefont {B{\"u}se}\ \emph {et~al.}(2015)\citenamefont
  {B{\"u}se}, \citenamefont {Tischler}, \citenamefont {Juan},\ and\
  \citenamefont {Molina-Terriza}}]{Buese2015}%
  \BibitemOpen
  \bibfield  {author} {\bibinfo {author} {\bibfnamefont {A.}~\bibnamefont
  {B{\"u}se}}, \bibinfo {author} {\bibfnamefont {N.}~\bibnamefont {Tischler}},
  \bibinfo {author} {\bibfnamefont {M.~J.}\ \bibnamefont {Juan}}, \ and\
  \bibinfo {author} {\bibfnamefont {G.}~\bibnamefont {Molina-Terriza}},\
  }\href@noop {} {\bibfield  {journal} {\bibinfo  {journal} {Journal of
  Optics}\ }\textbf {\bibinfo {volume} {17}},\ \bibinfo {pages} {065201}
  (\bibinfo {year} {2015})}\BibitemShut {NoStop}%
\bibitem [{\citenamefont {Kato}\ and\ \citenamefont
  {Takaoka}(2002)}]{Kato2002}%
  \BibitemOpen
  \bibfield  {author} {\bibinfo {author} {\bibfnamefont {K.}~\bibnamefont
  {Kato}}\ and\ \bibinfo {author} {\bibfnamefont {E.}~\bibnamefont {Takaoka}},\
  }\href {\doibase 10.1364/AO.41.005040} {\bibfield  {journal} {\bibinfo
  {journal} {Applied Optics}\ }\textbf {\bibinfo {volume} {41}},\ \bibinfo
  {pages} {5040} (\bibinfo {year} {2002})}\BibitemShut {NoStop}%
\bibitem [{\citenamefont {Pignatiello}\ \emph {et~al.}(2007)\citenamefont
  {Pignatiello}, \citenamefont {{De Rosa}}, \citenamefont {Ferraro},
  \citenamefont {Grilli}, \citenamefont {{De Natale}}, \citenamefont {Arie},\
  and\ \citenamefont {{De Nicola}}}]{Pignatiello2007}%
  \BibitemOpen
  \bibfield  {author} {\bibinfo {author} {\bibfnamefont {F.}~\bibnamefont
  {Pignatiello}}, \bibinfo {author} {\bibfnamefont {M.}~\bibnamefont {{De
  Rosa}}}, \bibinfo {author} {\bibfnamefont {P.}~\bibnamefont {Ferraro}},
  \bibinfo {author} {\bibfnamefont {S.}~\bibnamefont {Grilli}}, \bibinfo
  {author} {\bibfnamefont {P.}~\bibnamefont {{De Natale}}}, \bibinfo {author}
  {\bibfnamefont {A.}~\bibnamefont {Arie}}, \ and\ \bibinfo {author}
  {\bibfnamefont {S.}~\bibnamefont {{De Nicola}}},\ }\href {\doibase
  10.1016/j.optcom.2007.04.045} {\bibfield  {journal} {\bibinfo  {journal}
  {Optics Communications}\ }\textbf {\bibinfo {volume} {277}},\ \bibinfo
  {pages} {14} (\bibinfo {year} {2007})}\BibitemShut {NoStop}%
\bibitem [{\citenamefont {Ghosh}\ \emph {et~al.}()\citenamefont {Ghosh},
  \citenamefont {Molina-Terriza}, \citenamefont {Piro}, \citenamefont
  {Dubreuil}, \citenamefont {Torres},\ and\ \citenamefont
  {Eschner}}]{Ghosh2014}%
  \BibitemOpen
  \bibfield  {author} {\bibinfo {author} {\bibfnamefont {J.}~\bibnamefont
  {Ghosh}}, \bibinfo {author} {\bibfnamefont {G.}~\bibnamefont
  {Molina-Terriza}}, \bibinfo {author} {\bibfnamefont {N.}~\bibnamefont
  {Piro}}, \bibinfo {author} {\bibfnamefont {L.}~\bibnamefont {Dubreuil}},
  \bibinfo {author} {\bibfnamefont {J.~P.}\ \bibnamefont {Torres}}, \ and\
  \bibinfo {author} {\bibfnamefont {J.}~\bibnamefont {Eschner}},\ }in\
  \href@noop {} {\emph {\bibinfo {booktitle} {12th International Conference on
  Fiber Optics and Photonics}}}\ (\bibinfo  {publisher} {Kharagpur India, 2014
  (OSA)})\ p.\ \bibinfo {pages} {T3A.81}\BibitemShut {NoStop}%
\bibitem [{\citenamefont {Brambilla}\ \emph {et~al.}(2010)\citenamefont
  {Brambilla}, \citenamefont {Caspani}, \citenamefont {Lugiato},\ and\
  \citenamefont {Gatti}}]{Brambilla2010}%
  \BibitemOpen
  \bibfield  {author} {\bibinfo {author} {\bibfnamefont {E.}~\bibnamefont
  {Brambilla}}, \bibinfo {author} {\bibfnamefont {L.}~\bibnamefont {Caspani}},
  \bibinfo {author} {\bibfnamefont {L.~A.}\ \bibnamefont {Lugiato}}, \ and\
  \bibinfo {author} {\bibfnamefont {A.}~\bibnamefont {Gatti}},\ }\href
  {\doibase 10.1103/PhysRevA.82.013835} {\bibfield  {journal} {\bibinfo
  {journal} {Physical Review A}\ }\textbf {\bibinfo {volume} {82}},\ \bibinfo
  {pages} {013835} (\bibinfo {year} {2010})}\BibitemShut {NoStop}%
\bibitem [{\citenamefont {Osorio}\ \emph {et~al.}(2008)\citenamefont {Osorio},
  \citenamefont {Valencia},\ and\ \citenamefont {Torres}}]{Osorio2008}%
  \BibitemOpen
  \bibfield  {author} {\bibinfo {author} {\bibfnamefont {C.~I.}\ \bibnamefont
  {Osorio}}, \bibinfo {author} {\bibfnamefont {A.}~\bibnamefont {Valencia}}, \
  and\ \bibinfo {author} {\bibfnamefont {J.~P.}\ \bibnamefont {Torres}},\
  }\href {\doibase 10.1088/1367-2630/10/11/113012} {\bibfield  {journal}
  {\bibinfo  {journal} {New Journal of Physics}\ }\textbf {\bibinfo {volume}
  {10}},\ \bibinfo {pages} {113012} (\bibinfo {year} {2008})}\BibitemShut
  {NoStop}%
\bibitem [{\citenamefont {Gatti}\ \emph {et~al.}(2009)\citenamefont {Gatti},
  \citenamefont {Brambilla}, \citenamefont {Caspani}, \citenamefont
  {Jedrkiewicz},\ and\ \citenamefont {Lugiato}}]{Gatti2009}%
  \BibitemOpen
  \bibfield  {author} {\bibinfo {author} {\bibfnamefont {A.}~\bibnamefont
  {Gatti}}, \bibinfo {author} {\bibfnamefont {E.}~\bibnamefont {Brambilla}},
  \bibinfo {author} {\bibfnamefont {L.}~\bibnamefont {Caspani}}, \bibinfo
  {author} {\bibfnamefont {O.}~\bibnamefont {Jedrkiewicz}}, \ and\ \bibinfo
  {author} {\bibfnamefont {L.}~\bibnamefont {Lugiato}},\ }\href {\doibase
  10.1103/PhysRevLett.102.223601} {\bibfield  {journal} {\bibinfo  {journal}
  {Physical Review Letters}\ }\textbf {\bibinfo {volume} {102}},\ \bibinfo
  {pages} {223601} (\bibinfo {year} {2009})}\BibitemShut {NoStop}%
\end{thebibliography}%
\bibliographystyle{apsrev4-1}

\end{document}